# IN-SITU GROWN ERBIUM-DOPED DIELECTRIC NANOPARTICLES IN SILICA-BASED TRANSPARENT OPTICAL FIBERS

Bernard Dussardier, Wilfried Blanc, Valérie Mauroy, Michèle Ude, Stanislaw Trzesien, Franck Mady, Mourad Benadesselam, Gérard Monnom

Université de Nice-Sophia Antipolis, LPMC, CNRS UMR 6622, Parc Valrose, 06108 Nice cedex 2, FRANCE

bernard.dussardier@unice.fr

**Abstract:** An efficient method to fabricate transparent glass ceramic fibers containing in-situ grown $Er^{3+}$-doped oxide nanoparticles is presented. Characterization of the drawn fibers exhibits low loss (0.4 dB/m) and broadened $Er^{3+}$ emission spectrum. This attractive method offers new scopes for fiber amplifiers.

## 1. INTRODUCTION

New innovations in developing rare-earth (RE)-doped optical fibers for power amplifiers and lasers require continuous improvements in the fiber's spectroscopic properties (like gain and efficiency characteristics, spectral hole burning, photodarkening, …) besides reduction in device size and economical efficiency. In this context, transparent glass ceramic fibers are studied. RE-ions are embedded within small enough particles to induce acceptable scattering loss and of optimized composition and structure, different from the host glass. Through this route, one keeps the advantages of silica as host material, and spectroscopic properties can be engineered through the choice of the DNP chemical composition.

We propose a straightforward fabrication technique, based on the MCVD process, allowing to embed RE ions within *in-situ* grown dielectric nanoparticles (DNP) in silica-based performs. The original route is based on the spontaneous phase separation mechanism. One key advantage of this process is that DNPs are grown during the course of the fabrication process and there is no need (or associated risks) to manipulate nanoparticles by an operator. Further, the process takes advantage for high compositional control and purity typical of the MCVD technique. This paper deals with nanoparticles growth when alkaline earth ions (Mg, Ca and Sr) are incorporated. Influence on the spectroscopic properties of $Er^{3+}$ ions is also presented.

## 2. FIBER FABRICATION

Silica performs were first prepared by the usual MCVD process. Germania along with a small amount of phosphorous was added during perform fabrication to raise refractive index of the core and for ease in fabrication, respectively. Erbium and alkaline-earth ions (Mg, Ca and Sr) were incorporated through the well known doping solution technique during the perform preparation. A detailed description of the technique adapted for DNP growing was reported elsewhere [1].

## 3. RESULTS

When alkaline-earth ions were incorporated, nanoparticles were observed in fiber cores through scanning electron microscopy analyses. Mean size diameter depends strongly on the composition. It is around 100 nm in the CaO- and SrO-based DNP [2]. It decreases down to 40 nm in MgO-doped fibers. Then, at the wavelength of 1530 nm, losses were measured to be 0.4 dB/m only. This value is compatible with amplifier applications.

The spectroscopic properties of erbium ions were investigated through measurements of the fluorescence emission spectra and lifetime. A distinct broadening of the spectrum is observed by as much as ~50% for fiber containing a high content of Mg.

## 4. CONCLUSION

A method to fabricate nanostructured $Er^{3+}$-doped fibers entirely through the MCVD process is presented. Low-loss fibers and modified Er-spectroscopy are demonstrated. This concept has great potentials as possible solutions to nowadays issues in amplifying fibers.

## ACKNOWLEDGEMENT

This work was partly supported by CNRS (France). LPMC is a member of the GIS 'GRIFON' (http://www.unice.fr/GIS/).